\documentclass[epsf]{article}
\usepackage{amssymb}
\usepackage{amsmath}
\usepackage{graphicx}

\newcounter{saveeqn}
\newcommand{\alpheqn}{\setcounter{saveeqn}{\value{equation}}%
\stepcounter{saveeqn}\setcounter{equation}{0}%
\renewcommand{\theequation}{\mbox{\arabic{saveeqn}-\alph{equation}}}}
\newcommand{\reseteqn}{\setcounter{equation}{\value{saveeqn}}%
\renewcommand{\theequation}{\arabic{equation}}}

\newcounter{savesub}

\begin{document}

A STUDY OF THE GRAVITATIONAL WAVE PULSAR SIGNAL WITH ORBITAL AND SPINDOWN EFFECTS\\

S. R. Valluri$^{\ast\dagger}$, K. M. Rao$^{\ast}$, P.
Wiegert$^{\dagger}$, F. A. Chishtie$^{\ast}$

email: valluri@uwo.ca, krao@uwo.ca, pwiegert@uwo.ca,
fchishti@uwo.ca

Departments of $^{\ast}$Applied Mathematics, $^{\dagger}$Physics
\& Astronomy, University of Western Ontario, London, Canada

\begin{abstract}
In this work we present analytic and numerical treatments of the
gravitational wave signal from a pulsar which includes spindown.
We consider phase corrections to a received monochromatic signal
due to rotational and elliptical orbital motion of the Earth, as
well as perturbations due to Jupiter and the Moon. We discuss the
Fourier transform of such a signal, which is expressed in terms of
well known special functions and lends itself to a tractable
numerical analysis.
\end{abstract}

\section{Introduction}

The detection of gravitational waves (GW) from astrophysical
sources is one of the outstanding problems in experimental
gravitation today. Gravitational wave detectors like the LIGO,
VIRGO, LISA, TAMA 300, GEO 600 and AIGO are opening a new window
for the study of a great variety of nonlinear curvature phenomena.
Detection of GW necessitates sufficiently long observation periods
to attain adequate Signal to Noise ratio. The data analysis for
continuous GW, for example from rapidly spinning neutron stars, is
an important problem for ground based detectors that demands
analytic, computational and experimental ingenuity.

In recent works \cite{JVD96, CQG2002} we have implemented the
Fourier transform (FT) of the Doppler shifted GW signal from a
pulsar with the Plane Wave Expansion in Spherical Harmonics
(PWESH). It turns out that the consequent analysis of the Fourier
Transform (FT) of the GW signal from a pulsar has a very
interesting and convenient development in terms of the resulting
spherical Bessel, generalized hypergeometric, Gamma and Legendre
functions.  These works considered frequency modulation of a GW
signal due to rotational and circular orbital motions of the
detector on the Earth.  In the present analysis, rotational and
orbital eccentric motions of the Earth, as well as perturbations
due to Jupiter and the Moon, and pulsar spindown are considered.
This formalism has a nice analytic representation for the GW
signal in terms of the special functions above. The signal can
then be studied as a function of various parameters associated
with the GW pulsar signal, as well as the orbital and rotational
parameters of the Earth. A brief analysis considering the
parameters for the Lunar and Jovian perturbations is included. A
detailed analysis of the spindown and perturbation effects has
been recently done \cite{VCV2005}. The numerical analysis of this
analytical expression for the signal offers a challenge for
efficient and fast numerical and parallel computation.

In this work, we present a variation of a parametrized model of
pulsar spindown discussed by Brady et al. \cite{BC2000} and
Jaranowski et al. \cite{JK1999} for the GW frequency and the phase
measured at the ground based detector that includes the spindown
parameters. The various types of motion of the detector can be
naturally incorporated into this parametrized spindown model.
Effects of amplitude modulation (AM), though not considered in
this paper, can also be incorporated into this formalism without
great difficulty.  We discuss the Fourier transform of the GW
signal including the orbital and rotational corrections. In
addition, a numerical study of the effects of the various detector
motions on the phase of the GW signal and the dependence of the
received signal on detector position and pulsar direction have
been done. This might be relevant to GW detectors.

\section{Gravitational Wave Signal with Spindown Corrections}

Starting from a slightly modified form of the parameterization
given by Brady and Creighton (2000)\cite{BC2000}, we obtain the
following expression for the phase of a gravitational wave signal
\cite{VCV2005} :

\begin{equation}
\phi(\tau, \lambda) = \pi f_0 \left[ x + \tau_{min} \underset{k =
0}{\overset{\infty}{\sum}} \frac{T_k(z)}{k + 1} \hspace{3pt}
\left(\frac{x}{\tau_{min}}\right) ^{k
+ 1} \right]\\
\end{equation}

\noindent The $T_k(z)$ ($-1 \leq T_k(z) \leq 1, \hspace{10pt} k
\geq 0, \hspace{10pt} -1 \leq z \leq 1$) are the Chebyshev
polynomials, and $ x = t + \frac{\vec{r}}{c}\cdot\hat{n} $, where
$\vec{r}(t)$ is the position vector of the detector in the Solar
System Barycentre (SSB) frame, and $\hat{n}$ is a unit vector in
the direction of the pulsar. $\tau_{min}$ is the spindown age of
the pulsar in years, and $f_0$ is the GW frequency without
spindown. The series in this expression sums to

\begin{equation}
\underset{k = 0}{\overset{\infty}{\sum}} \frac{T_k(z)}{k + 1}
\left(\frac{x}{\tau_{min}}\right) ^{k + 1} = -\frac{1}{2} \left[
\ln\left(1 - 2 \left( \frac{x}{\tau_{min}} \right) z +
\left(\frac{x}{\tau_{min}}\right)^2\right) - 1 \right]
\end{equation}

We have plotted this expression as a function of $x$ for
$\tau_{min} = 40$ and $1000$ years for $z = -0.8, 0.7$ and $0.9$
(see Appendix, Figure 1).

The GW signal is now given by

\begin{equation}
e^{i\phi(t, \lambda)}= e^{\frac{i \pi f_0 \tau_{min}} {2}} e^{i
\pi f_0 x} \left[1 - \frac{ i\pi f_0 \tau_{min}}{2} \left( -2
\left(\frac{x}{\tau_{min}}\right)z +
\left(\frac{x}{\tau_{min}}\right)^2 \right)  + ... \right] .
\end{equation}\

We have used the binomial expansion inside the square brackets.
We define the Spindown Moment Integrals as follows:

\begin{equation}
\int_{0}^{T}e^{i\pi f_{0} \left(x + \frac{\tau_{min}}{2} \right)}
\cdot e^{-i2\pi ft}\cdot \left( \frac{x + \frac{\tau_{min}}{2}
}{\tau_{min}}\right)^{k}dt ,
\end{equation}

If we define

\begin{equation}
I_{generic}=\int_{0}^{T} e^{ \left[ i\pi f_{0}(x +
\frac{\tau_{min}}{2})- i 2\pi f t \right] }dt ,
\end{equation}
\noindent then the $k$th Spindown Moment Integral can be written
as

\begin{equation}
\frac{1}{\left(i \pi \tau_{min} \right)^k} \frac{\partial ^k
I_{generic}} {\partial f_0 ^k} .
\end{equation}

Thus the Fourier transform of the signal can be written in terms
of the partial derivatives with respect to $f_0$ of $I_{generic}$,
which can be evaluated analytically.

\section{Corrections to the Position Vector of the\\ Detector}

In this section we briefly outline the corrections to the position
vector $\vec{r}(t)$ to account for the Keplerian ellipse, Earth's
rotation, and the perturbations due to Jupiter and the Moon.

An expansion of the elliptical orbit of the Earth, to second-order
in eccentricity, leads to the following expressions for the $x$
and $y$ components of $\vec{R}_{orb}$, the vector specifying the
Earth's orbital position (the $z$ component is 0) \cite{MD1999} :

\alpheqn
\begin{equation}
x(t) = a\left(1-\frac{3}{8}e^2 \right) \cos M + \frac{ae}{2}\cos
2M + \frac{3}{8}ae^2 \cos 3M - \frac{3}{2}ae
\end{equation}

\begin{equation}
y(t) = a\left(1-\frac{3}{8}e^2 \right) \sin M + \frac{ae}{2}\sin
2M + \frac{3}{8}ae^2 \sin 3M - \frac{1}{4}ae^2 \sin M
\end{equation}
\reseteqn

\noindent ($a =$ Sun-Earth distance, $M = \omega_{orb} t$, $e =$
eccentricity of orbit.) If the radius of Jupiter's orbit is much
greater than that of the Earth, we can consider the Sun, to a
first approximation, as moving around the Sun-Jupiter (S-J)
barycentre, and the Earth orbiting the Sun. Then, the perturbation
$\vec{R}_{J}$ due to Jupiter is \cite{VCV2005} :

\begin{equation}
\vec{R}_{J} = \left[R_J \left(\cos\left(\omega_{J}t \right) - 1
\right), R_J \sin\left(\omega_{J}t \right), 0 \right]
\end{equation}

\hspace{-19pt} where $\omega_{J}$ is the orbital angular frequency
of Jupiter, and $R_J$ is the distance of the Sun from S-J
barycentre.

Similarly, the vector from the Earth-Moon barycentre to the Earth
is:

\begin{equation}
\vec{R}_{M} = \left[
\begin{array}{c c c}
R_{EM}\left(\cos\left(\omega_{M}t\right)-1\right)\\
R_{EM}\sin\left( \omega_{M}t \right) \\
0
\end{array} \right] .\\
\end{equation}

Here $\omega_{M}$ is the Moon's sidereal orbital angular
frequency, and $R_{EM}$ is the distance from Earth to the
Earth-Moon barycentre. Also, the vector specifying the detector's
position due to the rotating Earth, $\vec{R}_{rot}$, is given by

\begin{equation}
\vec{R}_{rot} = \left[
\begin{array}{c c c}
    R_E\sin\alpha\left(\cos\left(\omega_{rot} t\right) - 1 \right) \\
    R_E\sin\alpha\sin \left(\omega_{rot} t\right) \cos \varepsilon \\
    R_E\sin\alpha\sin \left((\omega_{rot} t\right) \sin \varepsilon \\
\end{array} \right] ,
\end{equation}

\noindent where $R_E =$ Earth's radius, $\alpha =$ detector
co-latitude, $\varepsilon =$ angular tilt of Earth's axis, and
$\omega_{rot} =$ Earth's sidereal rotational angular frequency.
The position of the detector is now given in the form:

\begin{equation}
\vec{r}(t) = \vec{R}_{orb}(t) + \vec{R}_{rot}(t) + \vec{R}_{J}(t)
+ \vec{R}_{M}(t)
\end{equation}

\section{Contributions of Perturbations to the Phase of the GW Signal}

The contribution of a perturbation to the phase of the GW signal
(in radians) is given by $\phi(t) = 2 \pi f_0 \frac{\vec{r}_i (t)
\cdot \hat{n}}{c}$, where $\vec{r}_i (t)$ is the contributing
vector for the perturbation \cite{CQG2002}. We have plotted the
phase contributions for the Earth's circular and elliptical
orbital and rotational perturbations, as well as contributions
from the Jovian and Lunar perturbations, as functions of $\theta$
(the angle of the pulsar from the orbital plane normal) and $\phi$
(the angle of the pulsar from the $x$-axis in the orbital plane),
for $t = 6$ lunar months ($6 \times 29.5$ d), $\alpha =
\frac{\pi}{2}$ and $f_0 = 1$ kHz (see Appendix, Figure 2.) In each
plot, the maximum phase shift occurs near $\theta = \frac{\pi}{2}$
(when the pulsar direction is in Earth's orbital plane). The
$\phi$-value at the maximum is more variable. The approximate
maximum phase shifts for the above perturbations are,
respectively, $6.3 \times 10^6$, $3.2 \times 10^6$, $1.8 \times
10^2$, $2.6 \times 10^4$ and $1.2 \times 10^2$ radians. For
comparison, the phase of the signal in this time period without
Doppler shifts ($2 \pi f_0 t$) is $9.6 \times 10^{10}$ radians.

We have also plotted the positional errors from using the circular
approximation versus those of equation 7, for a 1 kHz
gravitational wave (see Appendix, Figure 3). Four orders of
magnitude are gained, a substantial reduction in the error. In
addition, we see that the phase error can be quite large if we
ignore the effects of Jupiter and the Moon over long integration
times.

\section{The FT of the Perturbation-Corrected GW Pulsar Signal}

In Valluri et al. (2005) \cite{VCV2005}, we arrive at an analytic
expression for the Fourier transform of a GW signal, considering
the perturbations above:

\begin{equation}
\widetilde{h}(f) = \underset{r = -\infty }{\overset{\infty }{\sum
}}\underset{s = -\infty }{\overset{\infty }{\sum }} \underset{N =
-\infty }{\overset{\infty }{\sum }} \underset{n = -\infty
}{\overset{\infty }{\sum }} \underset{l = 0}{\overset{\infty
}{\sum }}\overset{l }{\underset{m = -l }{\sum }} \psi_0 \psi_1
\psi_2 \psi_3 \psi_4 \psi_5 \psi_6
\end{equation}

where \\

$ \psi_0(r,s,N,n,l,m,\alpha,\theta,\phi)= 4\pi i^{r+s+N+l}Y_{l
m}(\theta ,\phi )N_{l m}P_{l }^{m}(\cos \alpha)
e^{-i\left(r+s+N\right)\phi } \\$

$\psi_1(n,\theta, \phi,  f_0
)=T_{Er}\sqrt{\frac{\pi}{2}}e^{-i\frac{2\pi f_{0}a}{c}\sin \theta
\cos \phi -in\phi } i^{n}J_{n}\left( \frac{2\pi f_{0}}{c} a\sin
\theta \left( 1 -\frac{3}{8}e^{2} \right) \right) \\$

$\psi_2(l,n, m, r, s, N, f_0, \omega)=\left\{\frac{1-e^{i\pi (l
-B_{T})R}}{1-e^{i\pi (l -B_{T})}} \right\}
2e^{-iB_{T}\frac{\pi}{2}}\frac{1}{2^{2l +1}}
\\$

$\psi_3(k,l,m,n,r,s,N,f_0,\omega)=k^{l +\frac{1}{2}}\frac{\Gamma
\left(l +1\right) }{\Gamma \left(l +\frac{3}{2}\right)\Gamma
\left(\frac{l +B_{T}+2}{2}\right)\Gamma \left(\frac{l
-B_{T}+2}{2}\right)} \\$

$\psi_4(r,s,N,\theta,f_0)= J_{r}\left( \frac{2\pi f_{0}}{2c}ae\sin
\theta \right) J_{s}\left( \frac{2\pi f_{0}}{c}R_{J}\sin \theta
\right) J_{N}\left( \frac{2\pi f_{0}}{c}R_{EM}\sin \theta
\right)\\$

$\psi_5(\theta, \phi, f_0)= \exp\left[-\frac{i 2 \pi f_o}{c} \sin
\theta \cos \phi \left( \frac{3}{2} ae + R_J + R_{EM} \right)
\right]
\\$

$\psi_6(k,l,m,n,r,s,N,f_0,\omega)=_1F_{3}\left(l +1;l
+\frac{3}{2},\frac{l +B_{T}+2}{2},\frac{l
-B_{T}+2}{2};\frac{-k^{2}}{16} \right)
\\$

\noindent Here $c$ is the speed of light, and $\omega_0=2\pi f_0$. Also,\\
$ B_{T} = 2 \left( \frac{\omega -\omega_{0}}{\omega_{rot}}
+\frac{m}{2} +\frac{n}{2}\frac{\omega_{orb}}{\omega_{rot}}
+\frac{r}{2}\frac{2\omega_{orb}}{\omega_{rot}}
+\frac{s}{2}\frac{\omega_{J}}{\omega_{rot}}
+\frac{N}{2}\frac{\omega_{M}}{\omega_{rot}} \right)$ and
$k=\frac{4\pi f_0 R_E \sin(\alpha)}{c}$. \\

Feynman has thoroughly discussed the diffraction pattern due to
the factor in the curly brackets in the expression for $\psi_2 $
which gives the resultant amplitude due to $R$ equal oscillators
($R = 365$ in our case) \cite{Feynman}. This factor is analogous
to the expression for the scattering amplitude of an electron in a
crystal lattice.

The intensity of the signal, $\left|\tilde{h}(f) \right| ^2$, can
be expressed in analytic form from Equation 12. A detailed
numerical analysis of this expression for the variety of
parameters, which can take huge values $(> 50000)$, present in the
GW signal is a challenge for high performance parallel computation
\cite{CVRSW}.

\section{Conclusions}

We have presented in this paper the rudiments of a simple analysis
for spindown for sources of continuous GW. For the more
computationally intensive search over all sky positions and
spindown parameters, it is important to be able to calculate the
smallest number of independent parameter values which must be
sampled in order to cover the entire space of signals. The PWESH
improves the numerical accuracy and convergence of analytic FT's,
spindown corrections associated with the GW signal, and also
enables the estimation of parameters and distributions relevant to
GW Data Analysis. It has also found many important applications,
for example in the study of the multipole moments in the Cosmic
Microwave Background Anisotropies \cite{Peacock}. The proper blend
of analytic and numerical integration for accurate GW data
analysis remains an interesting question to be explored further.

\section{Acknowledgments}
We are deeply grateful to SHARCNET (Shared Hierarchical Academic
Research Cluster Network) and NSERC for grant support. We are also
greatly indebted to Drs. Nico Temme (CWI, Amsterdam), Martin Houde
(UWO), James Hilton(Astronomical Applications Dept., US Naval
Research Lab) and Manuel Gil (UWO) for valuable suggestions.

\clearpage
\section{Appendix}

\begin{figure}[h]
\centering
\includegraphics[width=12cm, height=2.535cm]{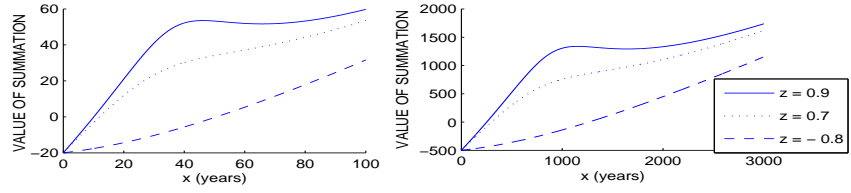}
\caption{Plot of summation as a function of $x$ for
$\tau_{min}=40$ years (left) and $\tau_{min}=1000$ years (right)
for different z values.}
\end{figure}

\begin{figure}[h]
\begin{minipage}[b]{0.3\textwidth}
\includegraphics[totalheight = 1.6 in]{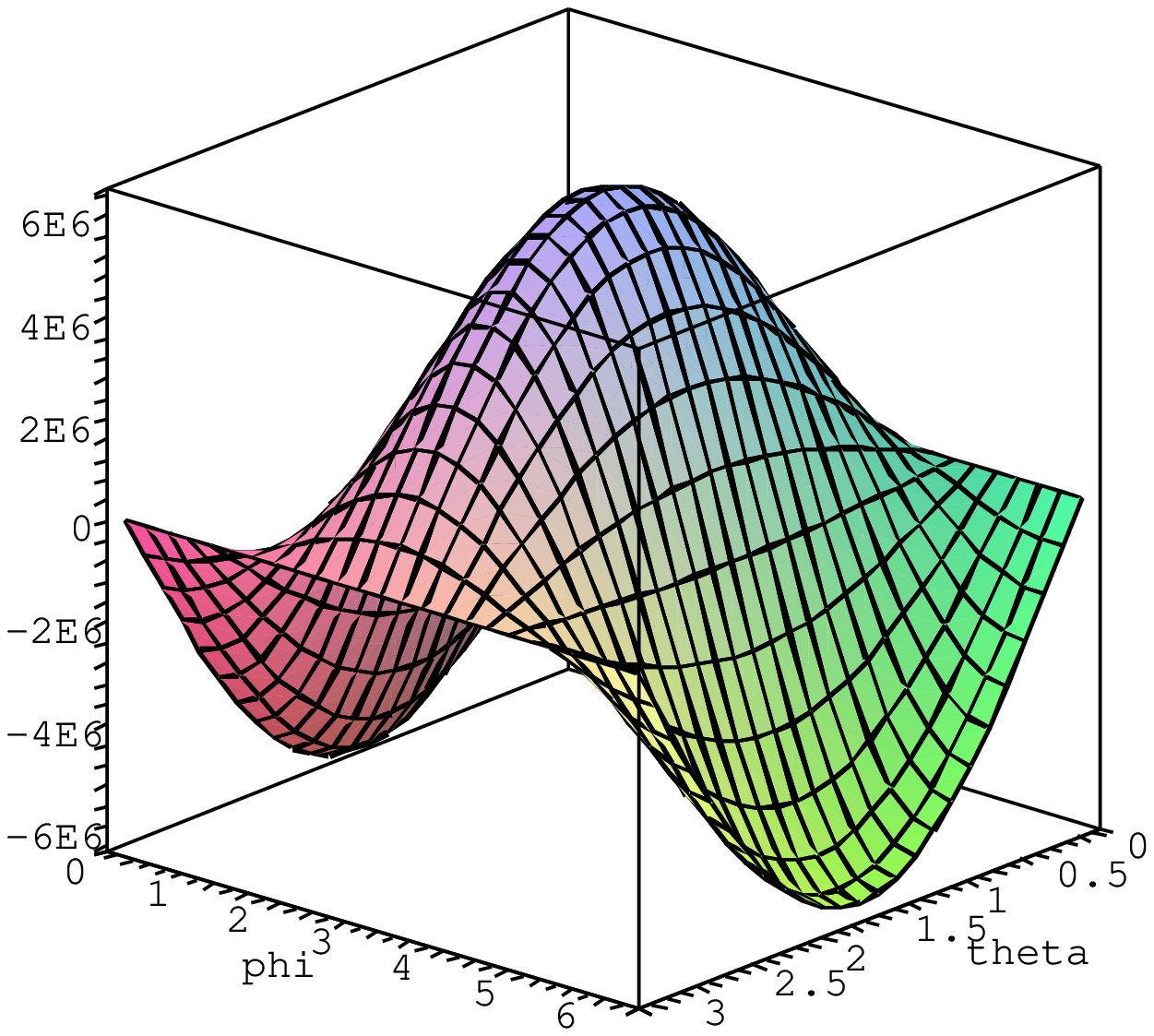}
\centering (a)
\end{minipage}
\begin{minipage}[b]{0.3\textwidth}
\includegraphics[totalheight = 1.6 in]{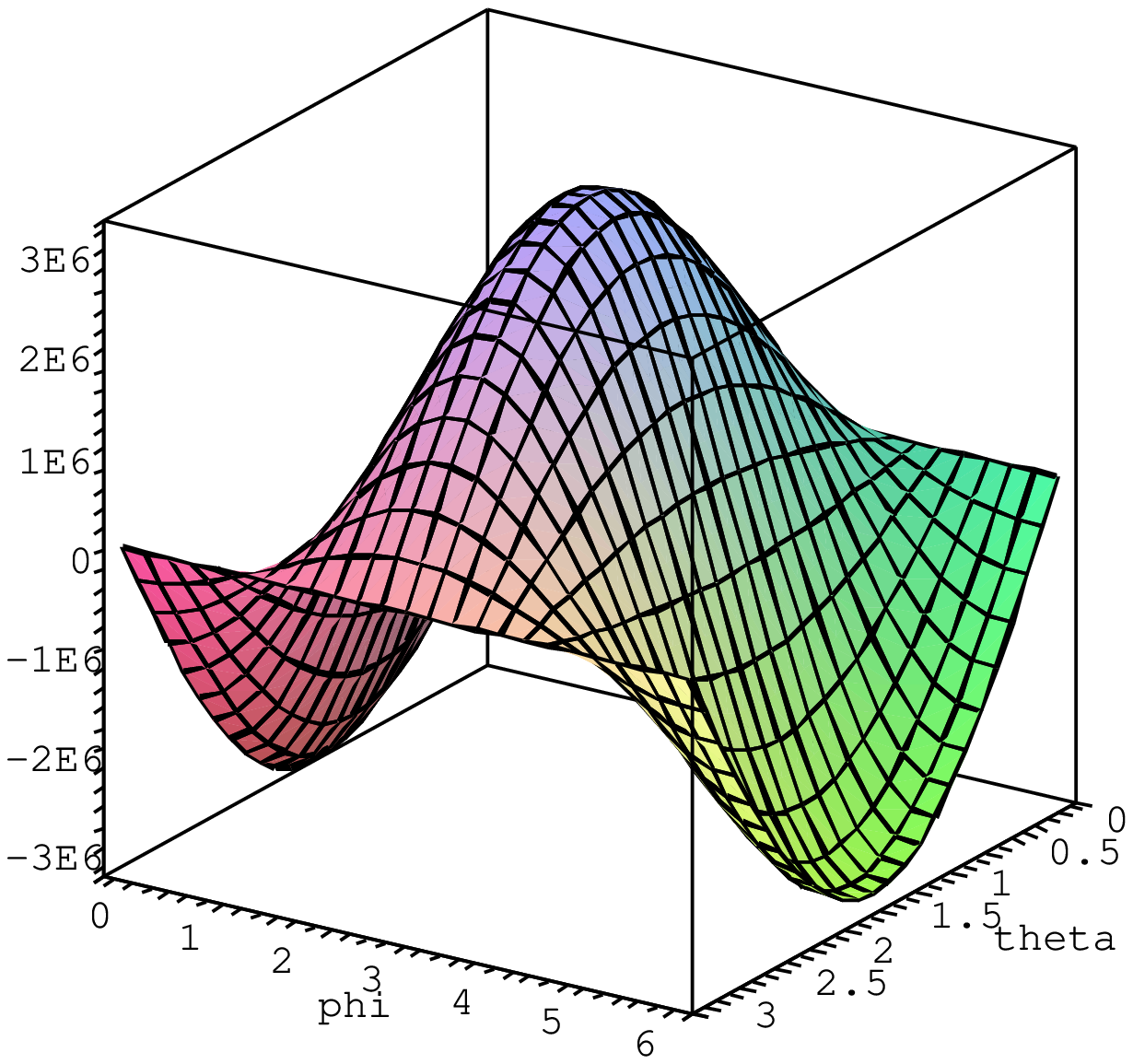}
\centering (b)
\end{minipage}
\begin{minipage}[b]{0.3\textwidth}
\includegraphics[totalheight = 1.6 in]{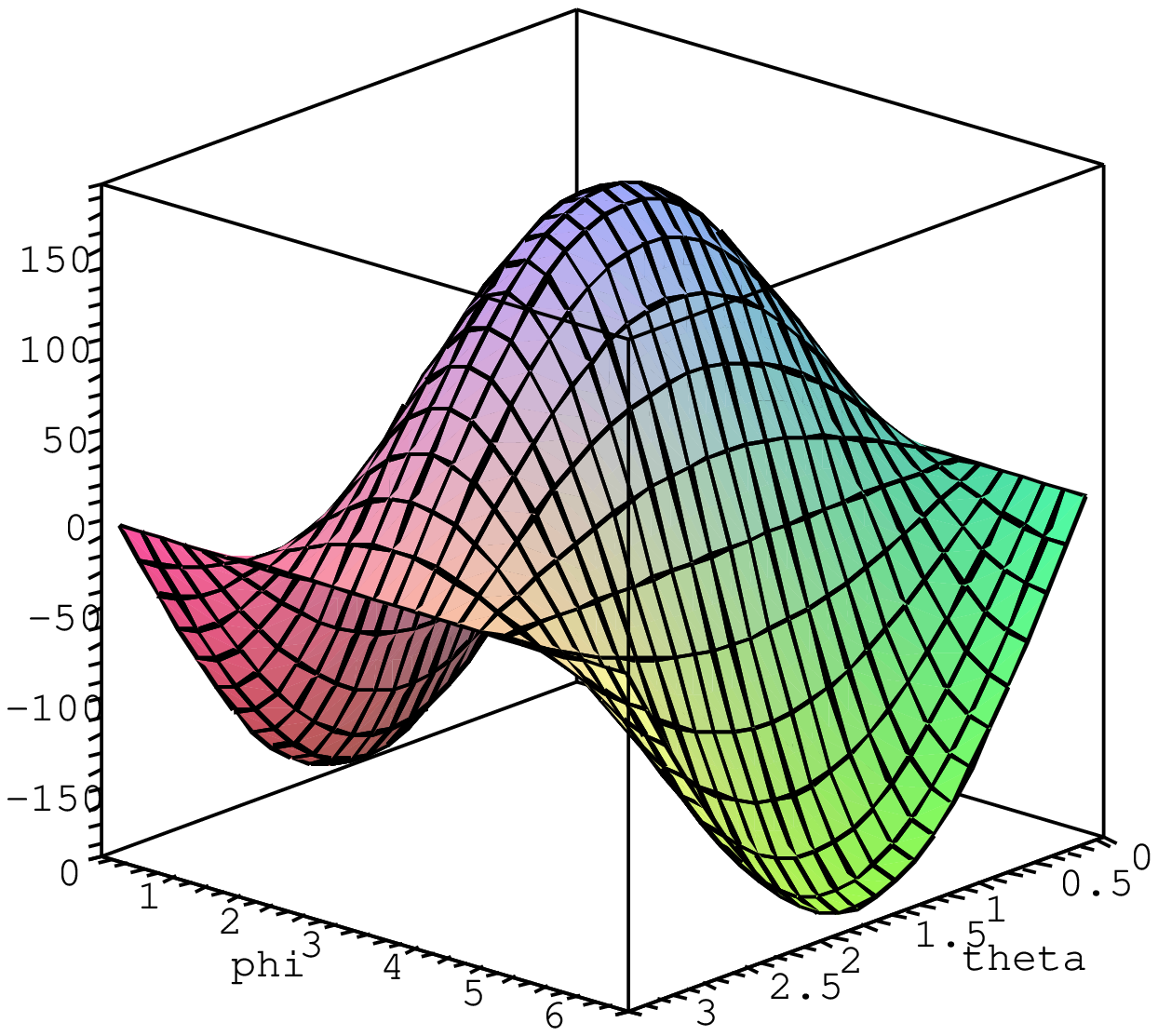}
\centering (c)
\end{minipage}
\begin{minipage}[b]{0.3\textwidth}
\includegraphics[totalheight = 1.6 in]{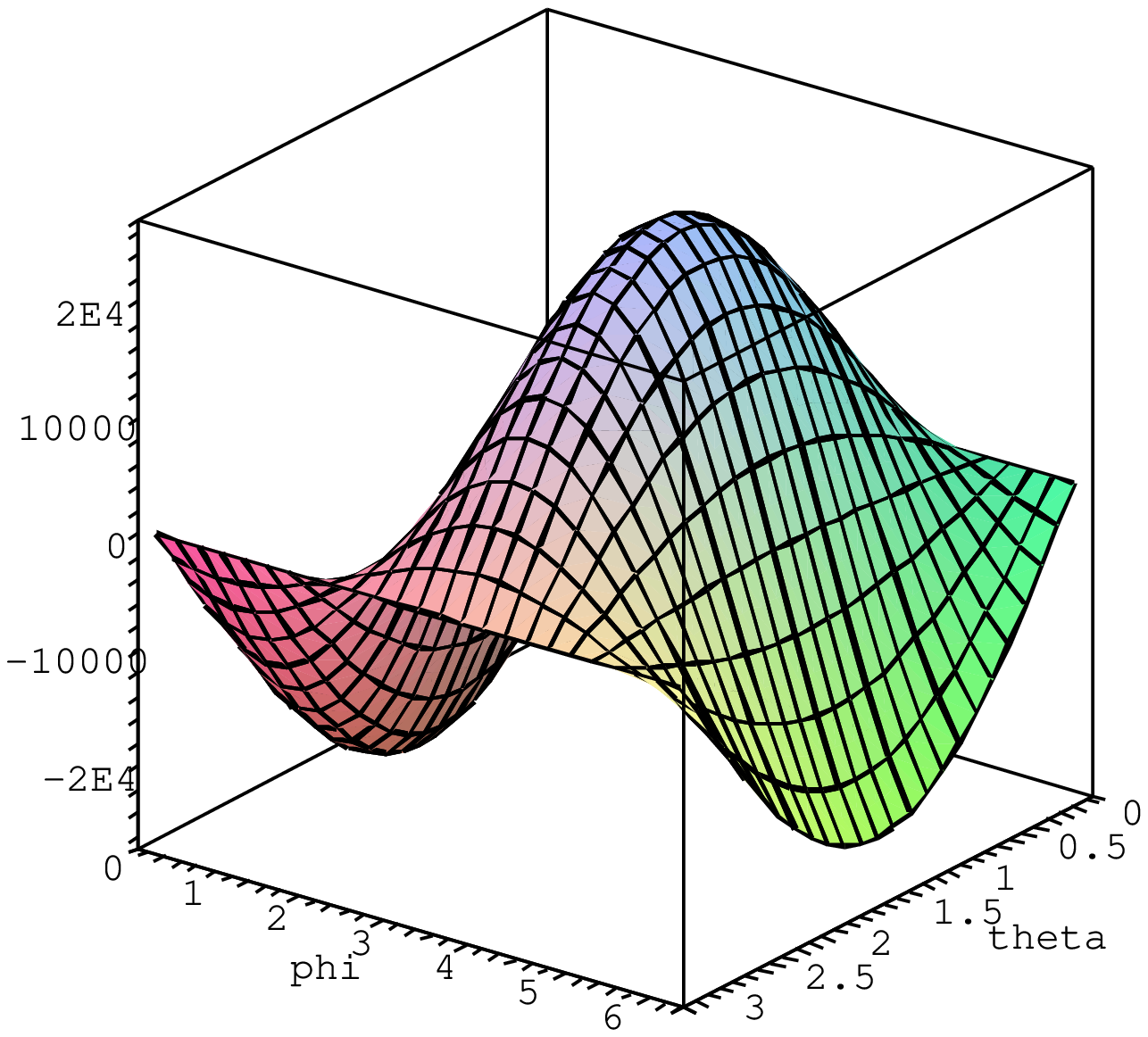}
\centering (d)
\end{minipage}
\hspace{25pt}
\begin{minipage}[b]{0.3\textwidth}
\includegraphics[totalheight = 1.6 in]{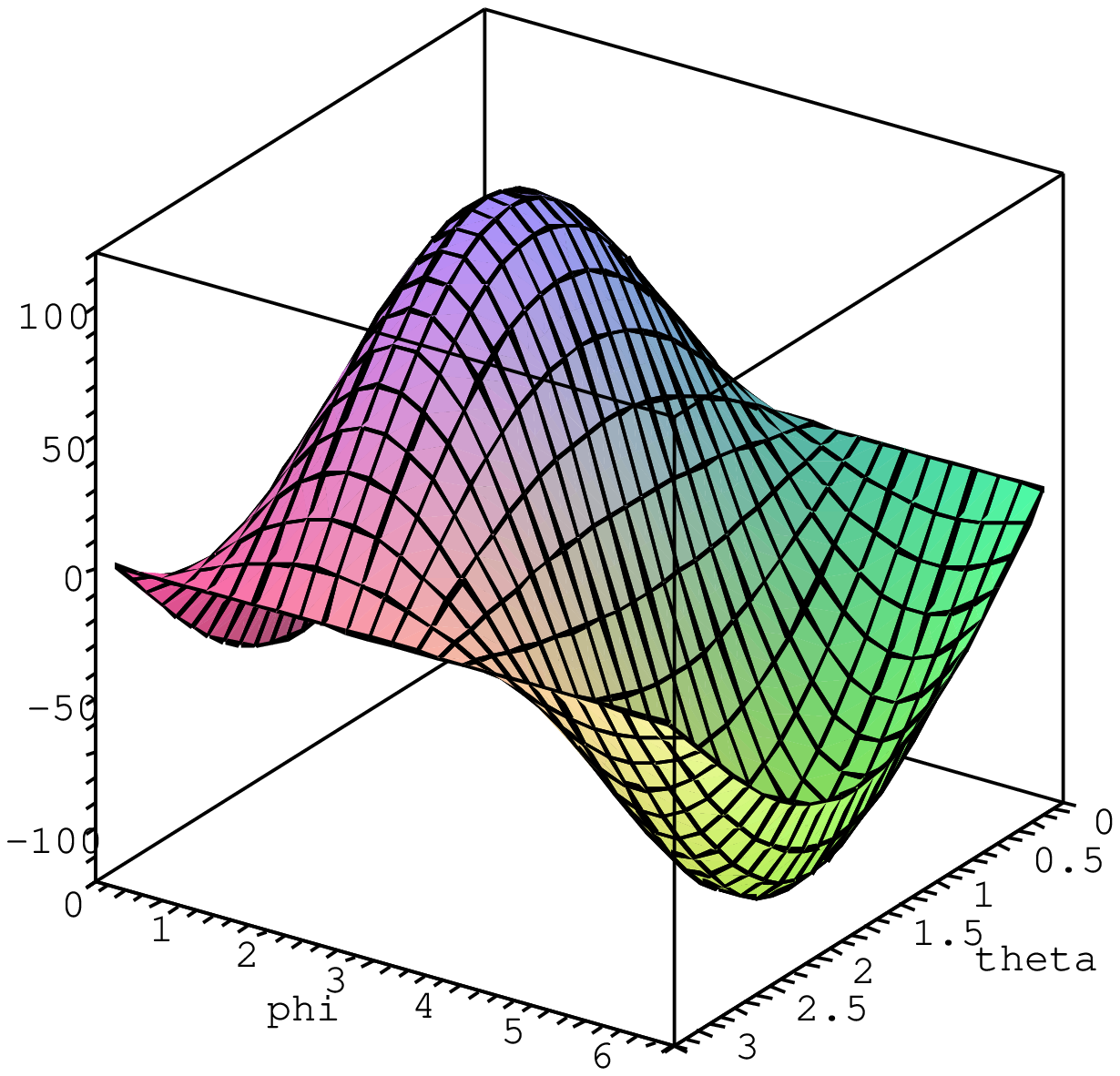}
\centering (e)
\end{minipage}
\caption{Plot of phase shift due to (a) circular orbit, (b)
elliptical orbit, (c) Earth's rotation, (d) Jupiter, and (e) the
Moon, as a function of $\theta$ and $\phi$.}
\end{figure}

\clearpage
\begin{figure}[h]
\centering
\includegraphics[scale=0.5]{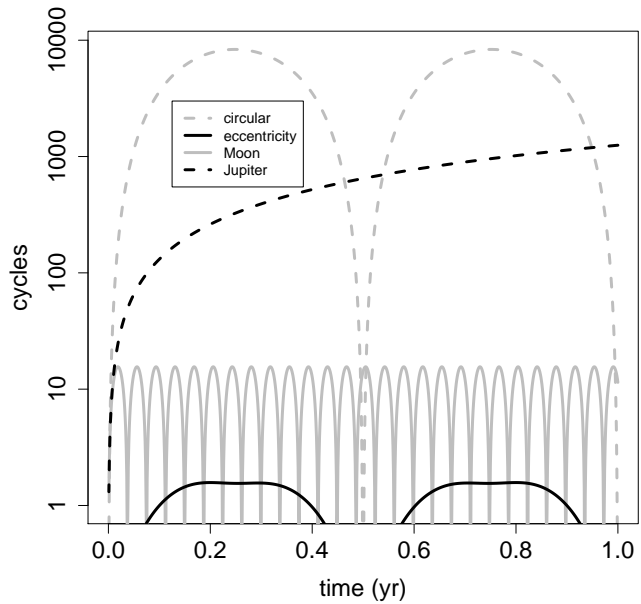}
\caption{The difference between the Earth's position as computed
by a circular approximation versus a Keplerian one, measured in
cycles of a 1 kHz gravitational wave over the course of one year
(dashed grey line). The (much reduced) difference when equation 7
is used is shown as a solid black line. Approximate contributions
to the motion of the Earth due to the Moon and Jupiter are shown
for comparison. }
\end{figure}

\end{document}